\documentclass[12pt]{article}
\usepackage{amssymb,tabularx,multirow,amsmath,natbib, epsfig, threeparttable, amstext}
\usepackage{setspace}
\usepackage{verbatim}
\usepackage{times}
\usepackage{fancyhdr}

\newcommand{\mbf}[1]{\mathbf{#1}}
\newcommand{\mbv}[1]{\mbox{\boldmath$#1$\unboldmath}}

\usepackage{setspace}
\RequirePackage[vmarginratio=1:2, outer=30mm, inner=26mm, lines=26]{geometry}
\usepackage{graphicx}
\usepackage{indentfirst}
\usepackage{caption}
\usepackage{subcaption}
\usepackage{natbib}
\usepackage{mathrsfs} 
\usepackage{floatrow}

\newcommand{\bes}{\begin{slide*}{}}
\newcommand{\es}{\end{slide*}}

\newcommand{\bitem}{\begin{itemize}}
\newcommand{\eitem}{\end{itemize}}
\newcommand{\blist}{\begin{list}{$\bullet$}{}}
\newcommand{\elist}{\end{list}}
\newcommand{\barray}{\begin{eqnarray*}}
\newcommand{\earray}{\end{eqnarray*}}
\newcommand{\bright}{\begin{flushright}}
\newcommand{\eright}{\end{flushright}}

\usepackage{booktabs}

\newcommand{\be}{\begin{eqnarray}}
\newcommand{\ee}{\end{eqnarray}}
\newcommand{\ba}{\begin{eqnarray*}}
\newcommand{\ea}{\end{eqnarray*}}

\def\boldfacefake #1{%
	\hbox{%
		\mathsurround=0pt
		\hbox to 0.25pt{$#1$\hss}%
		\hbox to 0.25pt{$#1$\hss}%
		\hbox {$#1$}%
	}%
}

\newcommand{\expect}{\mbox{\rm I\kern-.20em E}}
\newcommand{\reals}{\mbox{\rm I\kern-.20em R}}
\newcommand{\sreals}{\mbox{\small \rm I\kern-.20em R}}

\def\ba{\mathbf{a}}

\def\br{\mathbf{r}}
\def\bs{\mathbf{s}}

\def\bx{\mathbf{x}}
\def\by{\mathbf{y}}

\usepackage{color}
\usepackage{bm}
\usepackage{multirow}
\usepackage{amsmath}
\usepackage{amsfonts}
\usepackage{setspace}
\PassOptionsToPackage{round}{natbib}
\usepackage{natbib}
\usepackage{lineno}

\fancypagestyle{allpage}{
  \fancyhf{}
  \fancyfoot[C]{\ifnum\value{page}<0\relax\else\thepage\fi}
  \renewcommand{\headrulewidth}{0pt}
  \fancyhead[R]{P. L. MCDERMOTT AND C. K. WIKLE}
}
\usepackage{setspace}
\begin{document}
\bibliographystyle{environmetrics}

\doublespacing
\begin{titlepage}

\fancypagestyle{firstpage}{
  \fancyhf{}
   \fancyfoot[C]{\thepage}
  \renewcommand{\headrulewidth}{0pt}
  \fancyhead[L]{Research Article}
}

\title{A Model-Based Approach for Analog Spatio-Temporal Dynamic Forecasting}
\date{\vspace{-5ex}}
\author{Patrick L. McDermott\thanks{Correspondence to: P.L. McDermott, Department of Statistics, University of Missouri,  146 Middlebush Hall, Columbia, MO 65211 U.S.A.. E-mail: plmyt7@mail.missouri.edu} \footnotemark[2] ,\,\, Christopher K. Wikle\thanks{Department of Statistics, University of Missouri, Columbia, MO, 65211, U.S.A.}}

\maketitle
\thispagestyle{firstpage}
\begin{abstract}
Analog forecasting has been applied in a variety of fields for predicting future states of complex nonlinear systems that require flexible forecasting methods. Past analog methods have almost exclusively been used in an empirical framework without the structure of a model-based approach. We propose a Bayesian model framework for analog forecasting, building upon previous analog methods but accounting for parameter uncertainty.  Thus, unlike traditional analog forecasting methods, the use of Bayesian modeling allows one to rigorously quantify uncertainty to obtain realistic posterior predictive distributions. The model is applied to the long-lead time forecasting of mid-May averaged soil moisture anomalies in Iowa over a high-resolution grid of spatial locations. Sea Surface Temperature (SST) is used to find past time periods with similar trajectories to the current pre-forecast period. The analog model is developed on projection coefficients from a basis expansion of the soil moisture and SST fields.  Separate models are constructed for locations falling in each Iowa Crop Reporting District (CRD) and the forecasting ability of the proposed model is compared against a variety of alternative methods and metrics.
\end{abstract}
\textbf{Keywords:}
Bayesian, dynamical system, embedding vector, sea surface temperature, soil moisture
\baselineskip=22pt
\end{titlepage}
\setcounter{page}{2}
\pagestyle{allpage}


\section{Introduction}

Spatio-temporal processes govern a vast array of real-world phenomena. These processes are often dynamical in the sense that they can be viewed as multivariate spatial processes evolving in time, and are often nonlinear.  As the number and volume of datasets for such processes increase, there is growing interest in developing models that can be used to predict them. However, even in the context of linear dynamical spatio-temporal models (DSTMs), the curse of dimensionality has made the biggest implementation challenge the efficient reduction of dimensionality, either in state or parameter space \citep[e.g., see][for a recent overview]{wikle2015modern}.  These challenges are even more extreme for nonlinear DSTMs, although there are some general classes of parametric models for such processes that are designed for the typically quadratic nature of nonlinearity in these processes \citep[e.g.,][]{wikle2010general}.   However, in many cases, these models are computationally prohibitive, and thus, semi-parametric alternatives are of interest.  We investigate here the so-called class of {\it analog DSTMs.}

The origin of analog forecasting can be traced to the ``points of symmetry'' method of weather forecasting developed by \cite{weickmann1924} and championed by \cite{krick1941} and \cite{elliott1941}. The original idea is fairly simple and consists of a search through historical weather maps to find a ``match''  to the current map, and then assumes that the current forecast is just the realized future that was associated with the historical match.  Critical to this approach is having a sufficient ``library'' of historical cases and a mechanism to decide how best to evaluate a ÒmatchÓ (e.g., \cite{weickmann1924} suggested ``points of symmetry'' in time and space).  In the days before numerical weather prediction (i.e., numerical solutions to very high-dimensional systems of partial differential equations that govern the system behavior), these methods demonstrated some success.  For example, Krick is said to have used an analog technique to forecast successfully the weather expected with the D-Day invasion, although certain aspects of this story are somewhat controversial \citep{fleming2004sverre}. 

Though analog techniques remained known to weather forecasters \citep[e.g.,][]{gringorten1955methods}, they lost favor as numerical forecasts became increasingly prevalent with the advent of the digital computer.  Yet, \cite{lorenz1969atmospheric} considered analogs in evaluating short-term atmospheric predictability, linking the methods to the notions of chaos theory that he discovered \citep{lorenz1963deterministic}.  Although the method was shown to be less useful for short-term forecasting than the numerical weather prediction methods, it was eventually recognized as a viable empirical forecast method for medium to long-range forecasts \citep[e.g.,][]{barnett1978multifield,VDD} and for climate downscaling \citep[e.g.,][]{zorita1995stochastic,zorita1999analog}.   In general, these methods require very large libraries of potential analogs, although one can still effectively use the procedure with careful consideration of reduced portions of the state space \citep[e.g.][]{VDD}.

As mentioned above, traditional analog forecasting approaches seek to match of the current state of the system with some historical state, the idea being that the future would then evolve like the past.  One can formalize this more clearly in the context of dynamical systems. Chaotic dynamical systems are the result of nonlinear processes that are sensitive to initial conditions \citep[e.g.,][Chap. 3]{CandW2011}. Often dynamical systems will primarily inhabit a particular subset of the entire phase space as they move through time.  Informally, this subset of the phase space is referred to as an attractor. The goal of analog forecasting is to trace the path of the system along the attractor into the future given some initial conditions.
The popular and successful scalar analog forecasting algorithm of \cite{sug} was constructed as a simple solution based on this idea. 

As the method of \cite{sug} suggests, a key element of analog forecasting involves representing a dynamical system through state-space reconstruction methods.  For analog forecasting, state-space reconstruction is typically accomplished through the construction of embedding vectors in conjunction with the theory of \cite{Takens}. Takens' theorem establishes that one can recover the shape of the attractor, in situations where the initial conditions are not completely observed, through the construction of {\it embedding vectors}, which are just vectors of various multiples of a specified delay (i.e., lag) time \citep[e.g., see Chapter 3 of][for an overview]{CandW2011}.  Thus, the essence of analog forecasting involves creating so-called ``lagged embedding vectors'' from historical data and comparing those to an embedding vector constructed using the state at the current time (i.e., the {\it initial condition}). Then one assumes the system will continue to evolve in a manner similar to those historical embedding vectors that are ``closest'' to the initial embedding vector (i.e., analogs). As noted in  \cite{Zhao}, the embedding vectors have utility beyond the benefit provided in situations with partially observed initial conditions. That is, embedding vectors allow one to find analogs to the initial condition by comparing trajectories of the state process, instead of merely comparing one instant in time, and thus, embedding vectors capture dynamical characteristics of the system that are important when making forecasts. 
 
Finding analogs that have the same future trajectory as the current state is vital to the success of analog forecasting. The majority of previous analog forecasting methods have used Euclidean distance as a measure of similarity between embedding vectors and have often only considered univariate states. Since finding robust analogs depends on embedding vectors being spatially similar over time, it's not clear if Euclidean distance always leads to optimal analogs, especially for spatio-temporal state processes. The model presented here uses a Procrustes distance \citep[e.g.,][]{HTF} to identify analogs. 

In  \cite{sug}, the future values from the identified historical analogs are averaged to produce a forecast. One can think of this as finding the average of the k-nearest neighbors from the initial condition. Instead of averaging the nearest neighbors, others  \citep[e.g.,][]{Zhao} have used kernels to weight each neighbor.  A popular linear analog solution is the so-called constructed analogue (CA) method, developed in \cite{VDD} for climatological long-lead forecasting. As demonstrated in \cite{Tippett}, CA gives weights that imply the same forecast as linear regression and, thus, suffers from the same issues as linear regression in highly nonlinear settings. While CA can be a successful alternative for particular situations, it lacks the flexibility of many nonlinear methods.

Despite their operational success in meteorology and climatology, analog methods have not been considered directly in statistics.  Indeed, to our knowledge, none of the previous embedding-based analog forecasting methods have used a model-based framework that allowed for probability-based quantification of uncertainty in the produced forecasts. Here we combine some of the previous concepts that have been considered in analog forecasting (e.g., embeddings, non-Euclidean distances, nearest neighbors, and kernel weighting) and use them in a Bayesian framework. Many previous analog forecasting methods leave several choices up to the user, such as the length of the embedding vectors and the number of neighbors to consider or use simple heuristics to pick these vital parameters. The induced uncertainty associated with these choices has not typically been factored into the analog forecast uncertainty. 

 In nonlinear forecasting one can easily obtain spurious results by overfitting and care must be taken when evaluating the quality of prediction for a particular method.  Bayesian methods allow one to generate forecasts that are accompanied with a model-based measure of uncertainty.  We employ Markov Chain Monte Carlo (MCMC) methods to obtain parameter estimates and posterior predictive distributions for the forecasts. These simulation techniques lead to implicit model averaging, since forecasts are generated across the parameter space. A parameterized kernel is utilized to weight the selected historical analogs for forecasting. 
 
Long-lead forecasting in meteorology and oceanography refers to a forecast beyond the maximum skill range of deterministic numerical forecast models (i.e., about 10 days).  Because the ocean time-scale is much slower than the atmosphere, and there are known long-range relationships between ocean variability and the associated atmospheric response (i.e., so-called ``teleconnections''; \citet{philander1990}), it is often possible to obtain skillful forecasts of oceanic and atmospheric response variables at long lead times (sometimes as long as 12 months or so).  Traditionally, this is one area of weather forecasting where statistical models can do as well or better than deterministic models \citep[e.g.,][]{barnston1999predictive,jan2005did,barnston2012skill}.   We apply our Bayesian analog model to the problem of long-lead forecasting of a high-resolution spatial data set of soil moisture anomalies over the state of Iowa, U.S.  In particular, we use Pacific Sea Surface Temperature (SST) anomalies to find robust analogues for the current soil moisture. Empirical basis functions (univariate empirical orthogonal functions (EOFs), multivariate EOFs, and spatio-temporal canonical correlation analysis (CCA) basis functions)  are employed as spatial basis functions to reduce the dimension of the soil moisture and SST data and create embedding matrices.  We train the model on historical data to obtain parameter estimates and then forecast sequentially future soil moisture for a hold-out sample. These forecasts are compared to a variety of alternative long-lead forecasting models, with our Bayesian analog models showing the best or second best performance in 75\% of the cases considered.

We outline the motivating long-lead forecasting problem in Section 2.  This is followed in Section 3 by a description of the Bayesian analog forecasting methodology, illustrating dimension reduction, kernel weighting, choice of distance function, priors, and implementation. Section 4 gives the application results, followed in Section 5 by a brief summary and discussion of important issues associated with the analog DSTM approach.

\section{Motivating Problem}

Soil moisture is one of the critical components in hydrology, soil chemistry, and agriculture.  In the context of agricultural production of mid-lattitude crops such as wheat and maize, the amount of soil moisture at different phenomenological phases of crop growth can be an important indicator of potential yield and may suggest management mitigation strategies (e.g., irrigation and timing of planting and nutrient supplementation).  It has been shown that there is long-lead (several month) predictability of soil moisture in the U.S. given tropical Pacific sea surface temperature (SST) anomalies \citep[e.g.,][]{van2003performance}.  In turn, the El Ni\~no/La Ni\~na phenomena in the tropical Pacific ocean correspond to quasi-periodic anomalous warming and cooling, respectively, that can have dramatic effects on weather events across the globe \citep[e.g.,][]{philander1990}.  In the context of the long-lead forecasting of soil moisture, it is well accepted in climatology that the atmospheric teleconnective response (e.g., precipitation over North America) to Pacific SST variation is not linear \citep[e.g.,][]{hoerling1997nino}.  It is also well known that the soil response to precipitation and evapotranspiration is nonlinear \citep[e.g.,][]{guo2002temporal}. Thus, it is reasonable from a scientific perspective to consider nonlinear long-lead forecast models that include SST forcing on soil moisture.  Specifically, our interest here is to consider SST as a potential predictor of soil moisture across crop reporting districts (CRDs) in Iowa, U.S.,  at lead times of 1, 3, and 6 months so as to facilitate crop production management strategies.  The primary interest is with longer (6 month) lead times, given the importance of early planning for such agricultural management. 

The SST data come from the publicly available National Oceanic and Atmospheric Administration (NOAA) Extended Reconstruction Sea Surface Data (ERSST). We use a subset  of the ERSST data that covers a region with the coordinates  $30.5\,^{\circ}$S-$60.5\,^{\circ}$N latitude and $123.5 \,^{\circ}$E-$290.5 \,^{\circ}$E longitude. The SST data have a spatial resolution of $2\,^{\circ} \times 2,^{\circ}$ over 3,132 locations  (we exclude all locations that cover land; e.g., see Figure 1). The SST data are considered on a monthly time scale over a period from January 1915 through December 2008. We consider so-called ``anomalies,'' which consist of subtracting the monthly means developed from a climatological average for the period 1970-1999, as is common in the climate literature.  
\begin{figure}[H]
\centering
\captionsetup{font=small}
\includegraphics[width=18cm,height=14cm]{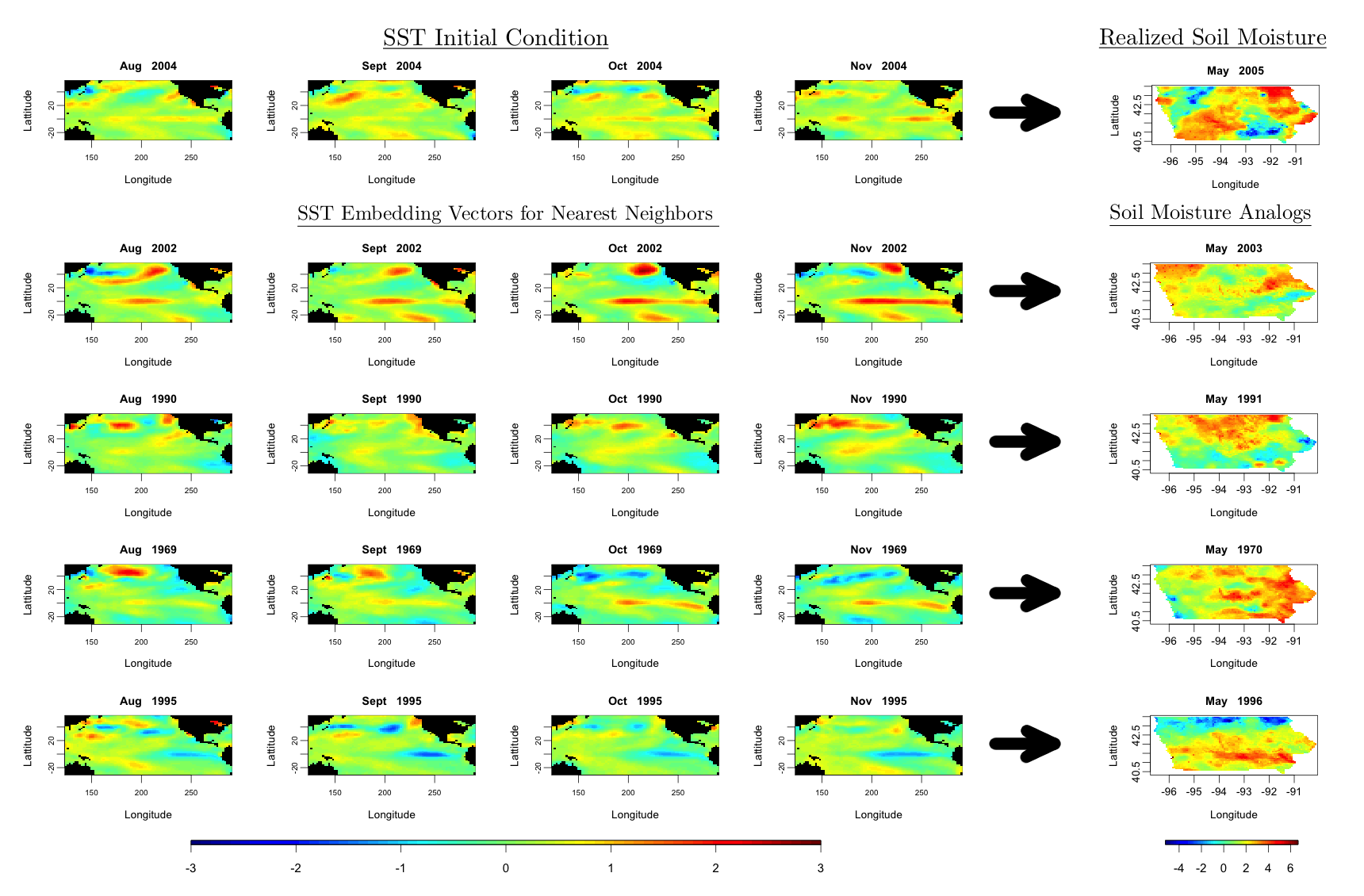}
\caption{ Illustrative example of using the analog approach to forecast 2005 mid-May averaged soil moisture anomalies in Iowa, U.S., with a lead time of 6 months using SST-based analogs. Displayed are the four nearest neighbors to the SST embedding sequence for the initial condition (starting at 6 months prior to May 2005). Each nearest neighbor SST embedding sequence corresponds to a past soil moisture realization that can be weighted to construct a forecast for the period of interest (May 2005). Note, how past SST embedding sequences (row 2 through 5) with a similar trajectory to the initial condition (top row), also have similar soil moisture anomaly fields six months in the future.  Embedding sequences are created with $h=1$ and $q=4$ in the notation described in Section 3. For illustration, Euclidean distance was used to find the displayed nearest neighbors. 
 }
\label{analog_ex}
\end{figure}

Given the importance of mid-May soil moisture to the planting of corn in Iowa, we seek to forecast the average soil moisture in an 11 day period centered on May 15.   The soil moisture data comes from a publicly available data set produced by the model described in \cite{smData}. In particular, the data are simulated from a so-called ``Variable Infiltration Capacity'' model. \cite{smData} show that the variability in the soil moisture model data is consistent with that of observed soil moisture. Spatially, the soil moisture model output data are gridded at the relatively high-resolution of $.0625\,^{\circ} \times .0625\,^{\circ}$. For the analysis presented here, we consider 4,053 spatial locations in Iowa contained in the 9 CRDs shown in Figure 2.  We model soil moisture in each CRD separately to facilitate model flexibility and to account for the known latitudinal and longitudinal gradients in climate variability over Iowa. Temporally, the data range from April and May 1915 -  April and May 2008 on a yearly time scale.  The soil moisture data were converted into anomalies by subtracting the climatological mean for the period from May 1970 - May 1999 for the May data (and similarly, for the April data) { (e.g., see Figure 1). Although we will predict mid-May anomalies, we use the mid-April information to supplement the choice of analogs (see below), with the justification that the patterns of spatial variation in soil moisture anomalies across this region is similar in April and May.} 
\begin{figure}[H]
\centering
\captionsetup{font=small}
\includegraphics[width=7.5cm,height=7.5cm]{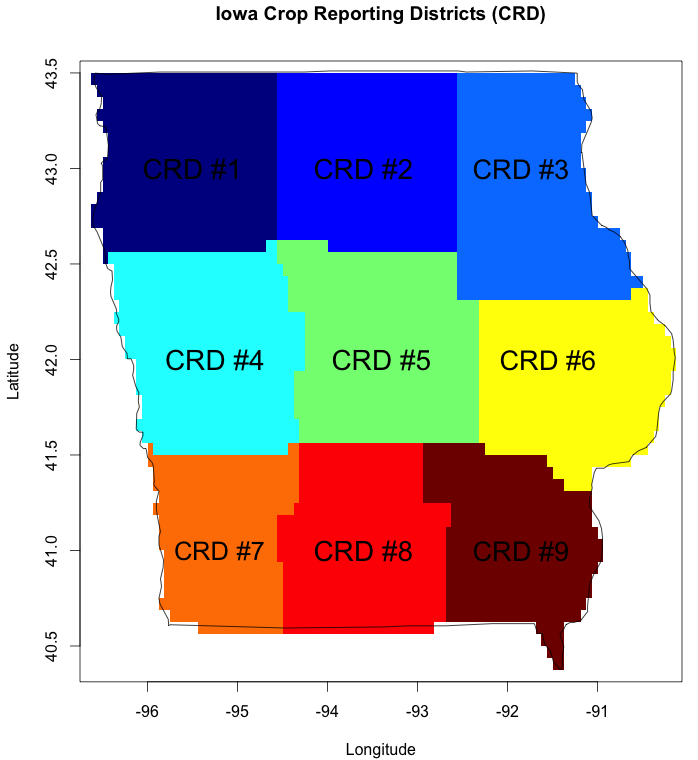}
\caption{The nine Crop Reporting Districts (CRDs) in Iowa, U.S., used to cluster locations when forecasting soil moisture. That is, a separate spatially-explicit forecast model is developed for each region.}  
\label{CRD_plot}
\end{figure}

\section{Bayesian Analog Forecasting Methodology}

There are a few practical issues that arise as one considers analog forecasting in more complicated spatio-temporal dynamical systems such as the SST/soil moisture problem outlined in Section 2.  In particular, we consider the situation where analogs for one portion of the state space (e.g., soil moisture) are developed from embedding vectors from another portion of the state space (e.g., SST).  We also consider the projection of high-dimensional spatial embedding vectors into a lower dimensional space, a nearest-neighbor kernel weighting approach, and a Procrustes based distance metric. Overall, we develop the spatio-temporal analog forecasting methodology in a model-based Bayesian framework, which to our knowledge has not been done before.  We refer to this Bayesian analog spatio-temporal model as a ``BA'' model. 

\subsection{Spatio-Temporal State Variables}

Suppose we want to forecast future values of a spatially referenced component of a dynamical system, say $y_{t}(\bs_i)$, for spatial location $\bs_i, i=1,\dots, n_y$ and times, $t=1,\dots,T,\dots$.  Define the $n_y$-vector $\by_t \equiv (y_t(\bs_1),\ldots,y_t(\bs_{n_y}))'$.  The analog prediction model developed here could be used to forecast these spatial vectors, but given the fact that $n_y$ is very large, we consider the projection of $\by_t$ onto a set of $p_\alpha$ spatial basis functions, $\{{\mbv \phi}_j, j=1,\ldots,p_\alpha\}$, where ${\mbv \phi}_j \equiv (\phi_j(\bs_1),\ldots,\phi_j(\bs_{n_y}))'$, and define the $n_y \times p_\alpha$ basis function matrix, ${\mbf \Phi} \equiv [{\mbv \phi}_1,\ldots,{\mbv \phi}_{p_\alpha}]$.   Then, the $p_\alpha$-dimensional projection coefficient vector at time $t$ is given by ${\mbv \alpha}_t =  ({\mbf \Phi}' {\mbf \Phi})^{-1} {\mbf \Phi}' \by_t$.  In practice, a wide variety of spatial dimension reduction basis vectors could be considered here \citep[e.g.,][]{wikle2010low}.  
We discuss particular choices for the spatial basis functions in Section 3.6 below.  


Now, consider a forcing (predictor) spatio-temporal vector given by ${\mbf x}_{t'} = (x_{t'}(\br_1),\ldots,x_{t'}(\br_{n_x}))'$, for spatial locations $\{\br_1,\ldots,\br_{n_x}\}$ and time $t'$.  Again, assuming that $n_x$ is large, we consider projections onto a reduced-rank set of spatial basis functions, say ${\mbf \Psi}  = [{\mbv \psi}_1,\ldots,{\mbv \psi}_{p_\beta}]$, where ${\mbv \psi}_k \equiv (\psi_k(\br_1),\ldots,\psi_k(\br_{n_x}))'$, $k=1,\ldots,p_{\beta}$.  
Specific choices for these basis functions are discussed in Section 3.6.  
In general, the $p_\beta$-dimensional projection coefficient vector at time $t'$ is given by ${\mbv \beta}_{t'} = ({\mbf \Psi}'{\mbf \Psi})^{-1}{\mbf \Psi}'\bx_{t'}$.  A special case occurs when the forcing vector $\bx_t$ can be lagged values of the response variable, $\by_t$, although this is not the case in our application.  

When working with embedding vectors constructed from basis function projections, we still want to take advantage of the benefits that come from a state-space reconstruction through embedding. We then construct {\it embedding matrices} by combining ${\mbv \beta}_{t'}$ for specified lags and number of vectors. Thus, we form the $p_\beta \times q$ embedding matrix for lag $h$ and with $q$ embedding vectors corresponding to time $t'$:
$$
{\mbf B}_{t'} = [{\mbv \beta}_{t'}, {\mbv \beta}_{t'-h},\ldots,{\mbv \beta}_{t'-h(q-1)}],
$$ 
for $t'=h(q-1)+1,\ldots,T'$, where  we require $q \ge 2$ and $h \ge 0$ (where both are integer valued).  The collection of these embedding matrices corresponds to our ``embedding library.'' Note, time $t'$ is assumed to correspond to time $t$ for the ${\mbf y}_t$ (i.e., ${\mbv \alpha}_t$) response process.  This notation is necessary because, in some situations, the forcing/predictor variables may be based on a different time resolution or might be inherently lagged relative to time $t$  {(e.g., see Figure 1)}.  

\subsection{Analog Forecasting Model}

The analog forecast model for predicting the $\alpha$-coefficients at time $t + \tau$ ($\tau = \{1,2,\ldots\}$) given observations up to time $t$ is given by 
\begin{equation}
{\mbv \alpha}_{t + \tau} = \sum_{\ell = { h(q-1) + 1}}^{t'-1}  w({\mbf B}_{t'},{\mbf B}_{\ell};{\mbv \theta}) {\mbv \alpha}_{\ell + \tau} + {\mbv \epsilon}_{t+\tau},
\label{eq:analogformod}
\end{equation}
where ${\mbv \epsilon}_{t + \tau} \sim N({\mbf 0},{\mbf \Lambda})$ and $w({\mbf B}_{t'},{\mbf B}_{\ell};{\mbv \theta})$ is a scalar weight based on the distance between embedding matrices, ${\mbf B}_{t'}$ and ${\mbf B}_{\ell}$ (see below), and is dependent on parameters, ${\mbv \theta}$.  Thus, the forecast of the response projection coefficients at time $t + \tau$ is a weighted combination of the past values of ${\mbv \alpha}_{\ell + \tau}$, where the weights are based on the distance between the embedding matrices for time $\ell$ and $t'$.  The idea is that if the past trajectory of the embedding vectors in the forcing projection coefficients is ``close'' to the recent trajectories corresponding to the time at which we want to forecast (sometimes called the ``initial condition''), then the associated past response $\tau$ lags beyond this time gets a high weight.  Correspondingly, if the distance between the embedding matrices is large, then that particular response vector gets a small weight.  After obtaining the prediction, $\hat{\mbv \alpha}_{t+\tau}$, the corresponding prediction in physical space is given by $\hat{\by}_{t+\tau} = {\mbf \Phi}\hat{\mbv \alpha}_{t+\tau}$.  Note, there is an implicit assumption here that the truncated components of the spatial basis expansion represent non-dynamical noise.  


\subsection{Kernel Weight Function}

The simplest analog approach corresponds to the case where the weight ($w(\cdot)$ in (\ref{eq:analogformod})) associated with the closest analog in our embedding library takes a value of 1 and all of the others get a weight of 0.  Alternatively, as shown in \cite{Tippett}, if one chooses the weights ``optimally'' based on minimizing the sum of squares, then this approach is equivalent to multivariate linear regression.  More generally, we use a kernel function to obtain the weights, such that the weights are larger the closer ${\mbf B}_{t'}$ is to ${\mbf B}_{\ell}$.  In particular, we use the radial Gaussian kernel, $w({\mbf E },{\mbf F};{\mbv \theta})=\text{exp}\{-d({\mbf E},{\mbf F};\theta_2)^2/{{2}\theta_1}\}$, where $d({\mbf E},{\mbf F};\theta_2)$ is some distance measure between ${\mbf E}$ and ${\mbf F}$ (see below) and the parameter $\theta_1$ controls the smoothness between neighboring analogs and can be thought of as a dispersion parameter. Larger values of $\theta_1$ will lead to more smoothing of the surrounding neighborhood and thus, will give forecasts with a lower variance. We note that there are many other suitable choices for the kernel function. For example, \cite{Zhao} and \cite{econGNP} use time dependent kernels, which potentially allow for incorporation of the evolving dynamic structure of the system. However, \cite{econGNP}  noted only marginal success when explicitly considering the dynamical structure in the kernel function. 

Now, suppose we denote the number of neighbors considered in a neighborhood in the vicinity of the embedding vector corresponding to the initial condition $({\mbf B}_{t'})$ as $m$. Thus, $m=1$ is the special case corresponding to the simple analog forecasting method mentioned above where only one embedding vector receives a weight of 1 and all of the others get a weight of 0. One can think of $m$ as defining the neighborhood around ${\mbf B}_{t'}$, i.e.,  ${\cal N}_m ({\mbf B}_{t'})$. By increasing $m$ and thus, the neighborhood size, the resulting predictions will be less variable but potentially more biased. In previous work, such as \cite{sug},  $m$ was specified to be $q+1$ since this is the minimum number of points needed to construct a bounding simplex. Instead of fixing $m$ at a particular value, we consider a range of values. Through the consideration of multiple neighborhood sizes we can achieve a model averaging effect that balances the effect of varying neighborhood sizes on predictions.  Finally, we note that by choosing the number of neighbors, we are effectively choosing a compactly supported kernel function:
 \begin{align}
 \label{kern}
\tilde{w}({\mbf B}_{t'},{\mbf B}_{\ell};{\mbv \theta};m)=
    \begin{cases}
  \  \text{exp}\left(\frac{-d({\mbf B}_{t'},{\mbf B}_{\ell};\theta_2)^2}{{2}\theta_1}\right)  \ , & \text{if}\ \   d({\mbf B}_{t'},{\mbf B}_{\ell};\theta_2)^2 \le h \\
      0, & \text{if}\ \  d({\mbf B}_{t'},{\mbf B}_{\ell};\theta_2)^2 >  h ,  \
    \end{cases}
  \end{align}
\\~\\
where $h=\displaystyle\max_{ {\ell} \in {\cal N}_m ({\mbf B}_{t'})} d({\mbf B}_{t'},{\mbf B}_{\ell};\theta_2)^2 $ and ${\cal N}_m ({\mbf B}_{t'})$ denotes the $m$ nearest neighbors to ${\mbf B}_{t'}$ according to the chosen distance metric. Thus, $h$ denotes the maximum distance in the neighborhood ${\cal N}_m ({\mbf B}_{t'})$ and $w({\mbf B}_{t'},{\mbf B}_{\ell};{\mbv \theta};m)$ is simply the normalized version of $\tilde{w}({\mbf B}_{t'},{\mbf B}_{\ell};{\mbv \theta};m)$, so as to force the weights to sum to one. As noted throughout the literature, the standard radial Gaussian kernel function has the property of positive definiteness. \cite{procPos} show that one can replace Euclidean distance with the Procrustes distance (see below) in the radial Gaussian kernel and still preserve positive definiteness. 

Many of the analog forecasting methods mentioned in Section 1 consider analog weights based on the process being predicted (i.e., $\{{\mbv \alpha}_t\}$ in our case). As mentioned, there is a strong scientific justification for the use of analogs based on the SST forcing (i.e., $\{{\mbv \beta}_t\}$) in the problem of interest here.  However, depending on the problem, one might seek to build analog weights from the forcing or the response coefficients.  This suggests a distance that represents a weighted combination of the two, such as
\begin{equation}
\gamma \ d({\boldsymbol  B}_{t'},{\boldsymbol B}_\ell; \theta_2) +(1-\gamma) d({\boldsymbol  \alpha}_{\tilde{t}},{\boldsymbol \alpha}_{\tilde{t}-1}; \theta_2),
\label{eq:weighted_d}
\end{equation}
where ${\boldsymbol  \alpha}_{\tilde{t}}$ is an embedding matrix and $\gamma\sim \text{Unif}(0,1)$. 

Finally, it is important to note that changing the embedding dimension, $q$, can change how similar the initial condition embedding matrix is to a given embedding matrix in the embedding library and thus, can affect the forecast. In much the same way that varying the neighborhood parameter $m$ effects the model, changing $q$ can lead to dramatically different forecasts.  Hence, in addition to a prior distribution on $m$, we consider a prior distribution on $q$ as well (see below).

\subsection{Procrustes Distance Metric}

When comparing the similarity between the embedding vectors or matrices associated with the initial condition and those associated with the embedding library, we would like to not only consider the similarity in an absolute distance sense, but also consider the similarity of their trajectories through phase space. Recall that one of the benefits of embedding the data is that for a given point we have more information than just the location of the state in phase space at a single point in time. Ideally, to find robust analogs we want to find past states that have a similar trajectory to the initial condition both in distance and trajectory ``shape.'' Having multi-dimensional embedding elements, we can take advantage of a Procrustes distance to find analogs (\cite{HTF}). 

In general, a Procrustes analysis tries to match a comparison shape to some target shape by finding optimal rotation, translation, and scaling between the two ``objects''.  This analysis is accomplished by transforming the comparison object to match up with the target object. In the analog forecasting context, the initial condition embedding matrix (${\mbf B}_{t'}$) is the target object and the other past embedding matrices ($\{{\mbf B}_{\ell}\}$) are the comparison objects. Once the comparison object has been transformed we can compare the distance between the target object and the transformed comparison object to get a measure of similarity. In particular, we calculate the Procrustes distance between ${\mbf B}_{t'}$ and ${\mbf B}_{\ell}$ using the scaling parameter $\theta_2$ and the orthonormal rotation matrix ${\mbf R}$ as follows (e.g., see \cite{HTF} for additional details):
\begin{equation}
d({\mbf B}_{t'},{\mbf B}_\ell; \theta_2)= \ ||{\mbf B}_{t'} - \hat{\theta}_2 \ {\mbf B}_\ell \  \hat{\mbf R}  ||_F ,
\label{eq:Pdist}
\end{equation}
where $\hat{\mbf R}= {\mbf U}{\mbf V}'$ comes from the singular value decomposition (SVD) given by $\tilde{\mbf B}_{t'} \tilde{\mbf B}'_{\ell} = {\mbf U}{\mbf D}{\mbf V}'$, where $\tilde{\mbf B}_{t'}$ and $\tilde{\mbf B}_{\ell}$ are column mean centered versions of ${\mbf B}_{t'}$ and ${\mbf B}_{\ell}$, respectively.   Finally, it can be shown that the solution to the positive scaling parameter is $\hat{\theta}_2=tr({\mbf D})/||{\mbf B}_{\ell} ||_F^2$, where the ``F'' subscript indicates a Frobenius matrix norm.  To scale the dissimilarity measure to a value between 0 and 1, (\ref{eq:Pdist}) is divided by $|| {\mbf B}_{\ell} - \boldsymbol{1} {\mbv \mu}^\prime ||_F $, where ${\mbv \mu} $ denotes the vector of column means of  ${\mbf B}_{\ell}$.  

As noted in \cite{dryden1998statistical}, the Procrustes distance is not symmetric.  There is some appeal to the lack of symmetry property here since the goal is to match past analogs to the initial condition, and the Procrustes distance allows us to examine transformations of all past analogs with respect to just the initial condition. The inverse relationship is not of interest since the initial condition is the only reference used to find analogs. Of course, other distance functions could be used here if desired -- the methodology is flexible to inclusion of alternatives.  In our analysis (summarized in Section 4), we found that using Procrustes distance to help select analogs improved prediction skill over using Euclidean distance.

\subsection{Bayesian Formulation}

To fully specify the Bayesian model, we assign prior distributions to the remaining parameters in the model. For the number of neighbors parameter, $m$, and the embedding matrix dimension parameter, $q$, we assign discrete uniform priors: $m \sim \text{DU}(m^{min}, m^{max})$ and $q \sim \text{DU}(q^{min}, q^{max})$, respectively. Bayesian k-nearest neighbors algorithms, such as \cite{chak}, have typically chosen to sample the number of neighbors with a discrete uniform prior.  

The kernel bandwidth parameter $\theta_1$ in (\ref{kern}) can be thought of as a variance parameter. As is often the case with variance parameters in Bayesian modeling, we consider an inverse-gamma prior:  $\theta_1 \sim \text{IG}( \alpha_{\theta_1},\beta_{\theta_1}) $. Alternatively, one could assign the bandwidth parameter a truncated normal (\cite{ktrunc}) or uniform prior (\cite{kuniform}). Although inverse-gamma priors are often chosen for their conjugacy with normal likelihoods, that is not the case here since the kernel function in (\ref{eq:analogformod}) will lead to a non-conjugate full-conditional distribution for $\theta_1$. Instead, the inverse gamma prior was chosen for its ability to accurately portray the skewed distribution of the bandwidth parameter. For the model variance, we let ${\mbf \Lambda} = \sigma^2_\epsilon {\mbf I}$ for simplicity, and assign an inverse-gamma prior, $\sigma_{\epsilon}^2 \sim \text{IG}(a, b)$.

\subsubsection{Implementation}
As described above, we create an embedding matrix with $h$ lag periods and $q$ embedding elements. Thus, the model could potentially be trained on every period from $h(q-1) + 1$ to $T^{tr}$ (where $T^{tr}$ is the user-specified upper end of the training period). Suppose we only want to train on a subset of all the possible training periods starting at period $t^{st}$, where $t^{st}\ge  h(q-1) + 1$. One might only train on a subset of all the training periods to either ease computational cost, account for temporal nonstationarity (e.g., climate change), or to help with mixing for the MCMC algorithm to follow.  We can then define this subset of training periods as $t^{tr}=t^{st},\dots,T^{tr}$. Since every period from $ h(q-1) + 1$ to $T^{tr}$ is a potential analog for a given initial condition from the training sequence $t^{tr}$, we use the following notation to denote the set of possible analogs for each training period with time index: $t^*=\{ h(q-1) + 1,\dots,t^{tr}-1,t^{tr}+1,\dots, T^{tr} \}$. Thus, $t^*$ helps index the library of embedding matrices at training period $t^{tr}$. Note that each period $t^{tr}$ in the subset of training periods is designated as the initial condition once, thus $t^*$ represents all the time periods from $h(q-1) + 1$ to $T^{tr}$ with the given initial condition removed. To ensure the number of training periods is the same for all values of $q$, we require $t^{st}\ge  h(q^{max}-1) + 1$, with $h(q^{max}-1) + 1$ always being the first period in $t^*$. 

Combining the likelihood from (\ref{eq:analogformod}) together with prior models outlined above, we get the following posterior distribution:
\begin{equation}
\label{post}
[\theta_1, \ m,  \ q, \ \sigma_{\epsilon}^2, \mid \{{\mbv \alpha}_t\}, \{{\mbv \beta}_{t'}\}] \propto \prod_{t^{tr}=t^{st}}^{T^{tr}}  [ {\mbv \alpha}_{t^{tr}} \mid \theta_1, \ m, \ q, \ \ \sigma_{\epsilon}^2, \  \{\mbv{\alpha}_{t^*}\},\{{\mbv \beta}_{t'}\}] \ [\theta_1] \ [m] \  [q] 
 \ [\sigma_{\epsilon}^2].
\end{equation}
To sample from the posterior in (\ref{post}) we use a standard Metropolis-within-Gibbs MCMC algorithm.


Forecasts for time periods outside the training period are obtained through Monte Carlo sampling from the posterior predictive distribution given the posterior draws from the parameters and the appropriate embedding matrices.

\subsection{Spatial Basis Functions}

As discussed previously, the methodology presented here is flexible in allowing one to select the spatial basis functions for the forcing (SST) and response (soil moisture) variables. It is common in atmospheric applications of analog forecasting to use empirical orthogonal functions (EOFs), which are just spatio-temporal principal components or, discrete solutions to a  Karhunen-Lo\`eve expansion,  (e.g., see \cite{CandW2011} for an overview).  In that case, ${\mbf \Phi}$ represents the eigenvectors of the decomposition of the empirical spatial covariance matrix of ${\mbf y}_t$ or, equivalently, as the singular value decomposition of the matrix ${\mbf Y} = [{\mbf y}_1,\ldots,{\mbf y}_T]$. Similarly, ${\mbf \Psi}$ represents a similar basis derived from the forcing variables. These EOF basis functions are orthonormal. Alternatively, given the dependence between the forcing and response variables, one could consider basis functions that factor in that dependence.  For example, one could consider spatial basis functions from so-called ``multivariate EOFs,'' which are derived from an estimate of the joint spatial covariance matrix, as well as basis functions derived from a spatio-temporal canonical correlation analysis (CCA) \cite[e.g., for details concerning the calculation of these, see][Chap. 5]{CandW2011}. We note that unlike the model with separate EOFs, the basis function matrix for both the multivariate EOFs and the CCA will no longer be orthonormal. In general, one could also consider nonlinear dimension reduction procedures, which could be particularly useful for the forcing variables that determine the analog weights.  Such dimension reduction procedures are not as useful for the response process because of the lack of a unique projection back to physical space.

\section{Long-Lead Soil Moisture Forecasting Application}
To evaluate the forecasting ability of the BA model described in Section 3, we train the model to forecast average mid-May soil moisture for the period from 1965 to 2001. All periods from 1915 through 2001 of average mid-May and mid-April are considered as potential analogs, even though the model is trained on a shorter period.  Note, the average mid-April soil moisture is only utilized when considering potential analogs; the April soil moisture anomalies have similar spatial patterns to the May anomalies. Once the model has been trained, the mid-May soil moisture for the seven years from 2002 to 2008 are forecasted at every location within each CRD. Seven years is a reasonable time span to assess the performance of the model without being concerned that the underlying structure of the system has changed fundamentally (e.g., climate change). The forecasting model is analyzed by comparing results against a number of different forecasting methods, along with multiple metrics as described in Section \ref{alt_methods} and \ref{error_met}. Lead times of ${\tau}=1,3,6$ months are evaluated, although primary interest is with the longest lead time (6 months) given its importance to agricultural management decision making. 
\subsection{Alternative Methods}
\label{alt_methods}
In climatological forecasting, one generally attempts to construct forecasting methods that will at least out-perform some n\"aive forecast. To our knowledge, the results presented here represent the most comprehensive model comparison published to date for testing the performance of analog forecasting methods.  The first model to which we compared was a multivariate multiple regression climate model (M1), where for each CRD, the projection coefficients of soil moisture for that district are regressed onto the lagged SST projection coefficients in order to get forecasts in the reduced rank space. These forecasts are then transformed back to physical space to get forecasts for each district. The aforementioned linear CA method (M2) with the same number of coefficients as the BA1 model (see below) is also compared.  In addition, we fit a first-order autoregressive model (M3), where the current (May) soil moisture is regressed onto the previous years soil moisture for a given location. In this setup, every soil moisture spatial location has an individual autoregression model.  In order to accommodate potential quasi-cyclic behavior present in the data, a second-order autoregressive model (M4) was also fit.  Further, n\"aive climatology and persistence forecasts are considered for comparison, as is customary in the long-lead forecasting literature. The climatology forecast (M5) consists of the yearly average of soil moisture in May. Two persistence forecasts are considered, one that uses the soil moisture from the previous May (M6) and another forecast that uses the soil moisture from April for a given forecasted year (M7), which is unrealistic in all but the 1-month lead time forecast scenario. Finally, we compare to a machine learning method by utilizing the ensemble learning random forest method \citep{HTF}. Similar to the model labeled M2, the random forest model (M8) forecasts the soil moisture projection coefficients for each district using SST projection coefficients. Each of the methods listed above are evaluated on the same hold-out sample as the BA models. 
\subsection{Evaluation Metrics}
\label{error_met}
With forecasts being produced over multiple time periods and hundreds of locations, there are a variety of ways to quantify the quality of the forecasts. For example, one could examine the mean squared errors (MSE) by CRD over all time periods or, alternatively, by time period over all locations. The results below use metrics average over time for each CRD. Since the BA model does not borrow information across districts, the model error should be assessed by district.  If we let $N$ represent the number of periods in the hold-out sample, then the MSE for each CRD is given by: 
\begin{equation}
\frac{\sum\limits_{t=T^{tr}}^{T^{tr}+N}  \ (\by_{t+\tau} -\hat{\by}_{t+\tau})^\prime(\by_{t+\tau} -\hat{\by}_{t+\tau})}{(N)(n_y)}.
\end{equation}


Throughout the literature, analog forecasting methods are evaluated based on  the correlation of their forecasted values with { realized} values.  Thus, we use the so-called ``Anomaly Correlation'' (AC) measure  \citep[e.g.,][]{wilks2006} to calculate the correlation between the {realized} and forecasted anomaly fields. This measure is especially useful as a way to compare the forecasted direction of an anomaly against the { realized} direction. The AC is similar to the Pearson product moment correlation, with sums over all locations and time periods in a district. In the literature, researchers have used both a centered and uncentered AC coefficient.  Similar to \cite{van2003performance}, we use the uncentered version:
\begin{equation}
\frac{ \sum\limits_{t=T^{tr}}^{T^{tr}+N}  \ \hat{\by}_{t+\tau}^\prime \by_{t+\tau} }{(\sum\limits_{t=T^{tr}}^{T^{tr}+N} \ \hat{\by}_{t+\tau}^\prime\hat{\by}_{t+\tau})(\sum\limits_{t=T^{tr}}^{T^{tr}+N} \ \by_{t+\tau}^\prime \by_{t+\tau})}.
\end{equation}

Many more metrics exist for comparing forecasts \citep[e.g.,][]{wilks2006}, but the two measures considered here are standard and allow us to sufficiently measure the distance and direction of forecasted values to  { realized} values. 

\subsection{Model Setup}
\label{mod_setup}

We consider four different BA models here.  The primary model described in Section 3 with separate univariate EOF basis functions for SST and soil moisture is referred to as BA1.  This same model, but with multivariate EOF basis functions is denoted BA2.  Similarly, the same model with CCA-derived basis functions is denoted BA3. Note that the spatio-temporal CCA was performed on reduced dimension space as described in Chapter 5 of \cite{CandW2011}.   Finally, 
the BA model with the weighted SST/soil moisture analog distances (i.e., equation \ref{eq:weighted_d}) with separate univariate EOF basis functions is denoted BA4.  

Finding past months with soil moistures that match the future month of interest relies on the success of the embedding matrices in identifying analogs (e.g., see Figure 1).While the number of periods in each embedding matrix is sampled, the lag and number of basis functions are both fixed. Both of these fixed parameters have a much smaller influence on the resulting embedding matrices than the parameters sampled in the BA model. For the BA1 model implementation presented here, the number of basis functions ($p_\beta$) for the embedding matrix is set to 12. Approximately 72 percent of the variation in the SST data is accounted for by these first 12 EOFs.  Similarly, the number of EOF coefficients for the soil moisture in each CRD was set to $p_\alpha = 5$.  Note, separate basis functions were estimated for each CRD. The BA4 model used the same setup in terms of number of basis functions as the BA1 model. Both the BA2 and BA3 models used 12 basis functions, which were estimated separately for each CRD. To complete the construction of the embedding matrices, the lag ($h$) is set to 1 for all of the lead times. The justification for using this lag value is that for relatively large values of $q$, $h=1$ will encompass subsets of vectors used for larger lag values, thus allowing one to sample over a wide range of historical values. Results indicated the model was more sensitive to different values of $q$ for higher $h$ values. 

Sampling of $m$ and $q$ was accomplished with the priors $m \sim \text{DU}(1,15)$ and $q \sim \text{DU}(2,24)$, respectively. The maximum for the number of neighbors parameter was limited to 15 to encourage smaller neighborhoods and thus, less bias. The computational cost associated with sampling over a larger range of $m$ seemed to significantly outweigh any benefit. Further, $q$ is sampled over a large enough set of values to allow embedding matrices to contain information as far back as two years. The bandwidth parameter is assigned the prior, $\theta_1 \sim \text{IG}(2,1)$. This weakly informative prior is used to promote smaller values of $\theta_1$. Since all of the distances are between 0 and 1, values of $\theta_1$ greater than one essentially weight all of the past values the same. Therefore, large values of $\theta_1$ all have a similar affect. The prior on the model variance is set such that, $\sigma_{\epsilon}^2 \sim \text{IG}(.001,.001)$. We note that the same priors were used for all 9 CRD models. For each model, the MCMC algorithm was run for $5,000$ iterations with the first $500$ iterations discarded as burn-in.

\subsection{Soil Moisture Results}
The results below are based on forecasts from the various models for the 7 year hold-out sample described above.  The results presented in Table \ref{table: tau_6_MSE}, show that in all the CRDs one of the BA models has the lowest MSE across all alternative models for a lead time of 6 months. From the perspective of a decision maker (say an agricultural producer), a lead time of 6 months is arguably more useful than a shorter lead time of 3 or 1 month. For the 6 month lead time, the other BA models perform slightly worse overall compared to the BA1 model. Not surprisingly, it appears that certain basis functions are more successful for certain districts.  This is consistent with other recent results that show varied performance for different spatial basis functions \citep{bradley2014comparison}. With a long enough hold-out sample one could potential learn which basis functions are most useful for finding analogs in a particular district. Viewing all of the lead times together, the nonlinear methods (BA models and M8) are most successful for 1 and 6 month lead times, while the most successful linear models (M1 and M2) perform the best for the 3 month lead time. We suspect that more nonlinearity is present in the scenarios were the BA and random forest models outperform the linear models. This is particularly true for the longest lead time (6 months), where it is believed that nonlinearity in the SST response is most prevalent.


Table  \ref{table: tau_6_AC} displays the ACs across the different methods for $\tau=6$. The lack of high correlation scores for any of the methods illustrates the difficulty involved with predicting the spatial pattern of soil moisture anomalies. For the 6 month lead time, we note that one of the BA models has the highest correlation coefficient  for 8 of the 9 districts.  Not surprisingly, the CA model (M2) performs relatively well for this metric. Often the primary metric used to evaluate CA forecasts, in works such as \cite{van2003performance}, is some type of correlation coefficient, not a loss function such as MSE.  The other two lead times show similar results to the MSE results for the BA models. Across all three lead times and all CRDs, one of the BA models had the lowest MSE in 14 of 27 cases and the highest AC in 14 of the 27 cases.  In 20 of 27 cases one of the BA models finished in either first or second in terms of lowest MSE. 

As mentioned previously, one of the benefits of the BA model, and using Bayesian modeling in general, is the ability to quantify uncertainty in forecasts. To illustrate this uncertainty, percentiles of the posterior predictive distribution are used to construct 95\% credible intervals over both space and time.  The  spatial map of CRD 6 in Figure 3 displays how the BA model performed for May 2003 with a lead time of 6 months. For this particular district we see that the model generally forecasts the correct direction and intensity of the soil moisture anomalies. 
The spatial plot in Figure 3, along with the time series plot to follow, both contain values that are standardized by their standard deviation within each district. This standardization is done across time for presentation purposes.
\begin{figure}[H]
\centering
\captionsetup{font=small}
\includegraphics[scale=0.35]{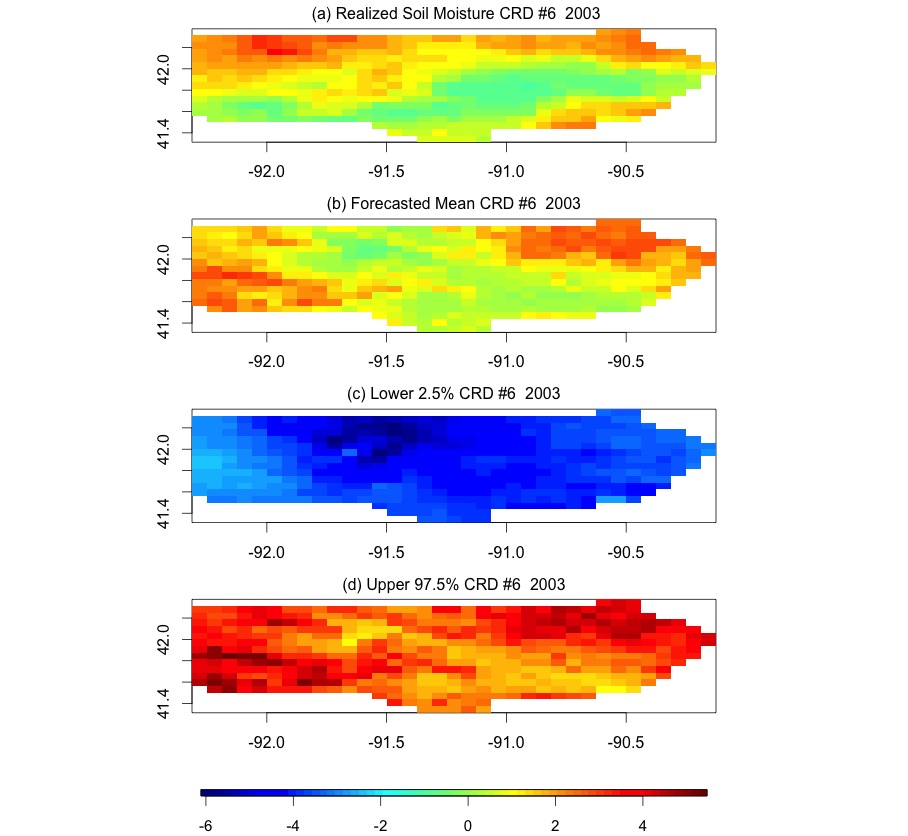}
\caption{(a) Realized soil moisture anomalies for CRD 6 for May 2003,(b) forecasted mean from the posterior predictive distribution,(c) lower 2.5th percentile from the posterior predictive distribution, and (d) upper 2.5th percentile from the posterior predictive distribution.}
\label{spat_map}
\end{figure}

The uncertainty in the forecasted soil moisture over time is presented in Figure 4 for a lead time of 6 months across all districts. For a particular year, the forecasts across all locations in a district were averaged to get the time series presented in Figure 4.  The model captures the correct trend over time for most of the CRDs, although there are notable exceptions -- some of the intervals are highly skewed and contain wide intervals.  We see that the uncertainty varies across time with some years having more narrow credible intervals than others. 
\begin{figure}[H]
\centering
\captionsetup{font=small}
\includegraphics[scale=0.35]{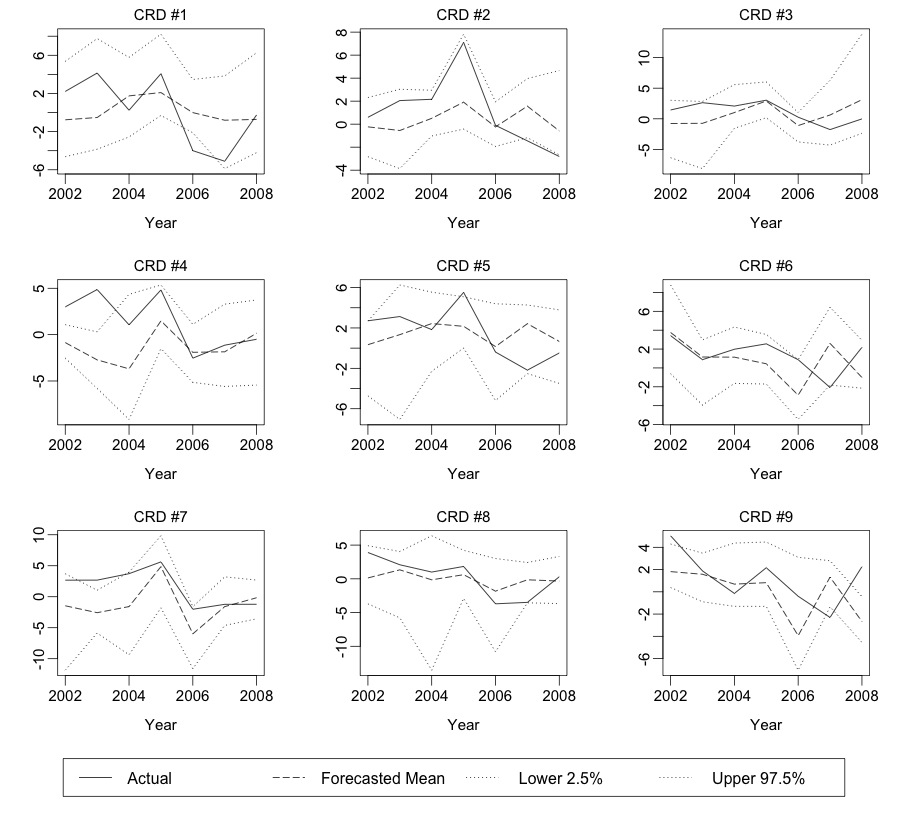}
\caption{Realized and 6 month lead time forecasted soil moisture for all 9 CRDs across the 7 years in the hold-out sample, along with 95\% confidence bands.  }
\label{temp_plot}
\end{figure}
\section{Discussion}
The long-lead prediction of high-dimensional nonlinear dynamic spatio-temporal processes is one of the most challenging problems in environmental statistics.  Outside of statistics, the analog forecasting approach has long been a useful nonparametric method for such problems due to its simplicity and for its implicit accounting for nonlinear dynamics.  The model we present here is an attempt to bring this approach into the statistics literature and to quantify the uncertainty associated with many of the model choices by placing it in a Bayesian framework.  As such, it is quite tractable for even high-dimensional problems due to its parsimony in parameter space. More importantly,  this Bayesian analog DSTM accomplishes the dual goal of quantifying the uncertainty present in analog forecasting, while also consistently producing high quality forecasts relative to competitors.  In addition to casting the approach in a formal Bayesian context, we incorporate novel components (e.g., Prucrustes-based distance metrics) and bring together disparate components from the analog literature (e.g., embedding vectors, kernel weighting, nearest neighbors, and basis-function dimension reduction).  


We applied our model to forecasting mid-May soil moisture over Iowa given embedding matrices derived from Pacific SST anomalies at various time lags.  Compared to eight alternative models, the BA model was the overall best performer in out-of-sample forecasts, both in terms of MSE and in terms of anomaly correlation metrics.  The BA model consistently forecasted the soil moisture trend and demonstrated that forecast skill is spatially and temporally dependent. 

There is much work to be done with regards to using analog forecasting in a model framework, and we hope that its introduction into the statistics literature will provide a basis for new research in nonlinear DSTMs. For example, there is still a need to strategically incorporate the dynamic structure into the kernel function. Using a different kernel may incorporate more parameters, which could easily be handled within the presented framework. Covariates that help explain some of the variation in the response are certainly another extension that could be considered within this framework. In addition, allowing for matrix weights in (\ref{eq:analogformod}), accounting for the dimension reduction truncation, and linking spatial regions (e.g., CRDs) through a fully hierarchical model, are obvious extensions. Overall, this is a promising direction for spatio-temporal modeling in environmental statistics. 

\section*{Acknowledgments}
This work was partially supported by the U.S. National Science Foundation (NSF) under grant DMS-104903. We would also like to thank an anonymous reviewer for helpful comments on an early draft.

\newpage
\bibliography{reference}
\newpage

\section*{Appendix: Soil Moisture Results}
\begin{table}[H]
\footnotesize
\captionsetup{font=footnotesize}
\centering
\begin{tabular}{p{1cm} p{1cm} p{1cm}  p{1cm}   p{1cm}   p{1cm}   p{1cm}  p{1cm}   p{1cm} p{1cm}  p{1cm}   p{1cm}  p{1cm} }
&  & & & &   \multicolumn{3}{c}{ Model}  \\
\hline
CRD & BA1 & BA2 &BA3 & BA4 & M1 & M2 & M3 & M4 & M5 & M6 & M7  & M8  \\ [1ex] 
\hline
 \#  1 &  4.854  &  7.558  & 6.749  & \bf{4.760}  &   7.79  &  8.11  & 5.121  & 5.153   & 5.139   & 9.425   &   5.51   & 6.624  \\
  \#  2 &  4.067  & 5.715  & 5.128  & \bf{3.964}  &  5.753  & 5.991  & 4.788  & 4.771   & 4.488   & 4.341   &  8.261   & 4.343  \\
  \#  3 &  \bf{5.311}  & 7.671  & 9.860 & 5.695 &  7.277  & 7.484  & 6.998  & 7.351   & 5.964   & 5.575   & 13.379   & 6.993  \\
  \#  4 &  5.013  & 4.345  & \bf{3.741}  & 4.877   &  5.975  & 6.305  & 5.214  & 5.279   &  4.88   & 6.706   &  7.433   & 4.697  \\
  \#  5 &  4.325  & 5.169 & \bf{3.897}  & 5.347  &  5.693  & 6.013  & 5.715  & 6.163   & 4.823   & 5.035   &  8.155   &   4.3  \\
  \#  6 &  5.943  & \bf{4.832} & 8.456  &  5.843 &  6.355  & 6.525  & 6.617  & 6.655   & 5.859   & 8.826   & 13.646   & 6.232  \\
  \#  7 &  \bf{3.372}  & 4.514  & 4.871 & 3.991 &  4.036  & 4.252  & 4.523  &  4.69   & 4.041   & 3.802   & 10.431   &  3.75  \\
  \#  8 &  \bf{4.855}  & 5.361  & 8.690 & 6.189   &  6.082  & 6.471  &  6.67  & 6.865   & 5.626   & 6.427   & 10.086   & 5.888  \\
  \#  9 &  3.496  & 6.448  & 4.423  & \bf{2.996}  & 4.547  &   4.8  & 4.627  & 4.588   &  3.67   & 5.879   &  7.595   & 5.614  \\
\hline

\end{tabular}
\caption{MSE values for a lead time of $\tau=6$, averaged over all periods in the hold-out sample, by CRD. For each CRD, the model with the lowest MSE is bolded. See Section \ref{alt_methods} for descriptions of each model listed. }
\label{table: tau_6_MSE}
\end{table}  
\begin{table}[H]
\footnotesize
\captionsetup{font=footnotesize}
\centering
\begin{tabular}{p{1cm} p{1cm} p{1cm}  p{1cm}   p{1cm}   p{1cm}   p{1cm}  p{1cm}   p{1cm} p{1cm}  p{1cm}   p{1cm}  p{1cm} }
&  & & & &   \multicolumn{3}{c}{ Model}  \\
\hline
CRD & BA1 & BA2 &BA3 & BA4 & M1 & M2 & M3 & M4 & M5 & M6 & M7  & M8  \\ [1ex] 
\hline
 \#  1 &   0.299 &   -0.272   &  -0.146  & \bf{0.419}   &  0.025  & -0.234  &  0.059  &  0.045   &  0.082   & -0.026   &  0.056   & 0.056  \\
\#  2 &   0.251  & -0.202   & -0.221  & \bf{0.320}  & 0.161  &  0.161  & -0.184  & -0.126   & -0.145   &  0.122   &  0.086   & 0.177  \\
\#  3 &   \bf{0.306}  & -0.420  & -0.486    & 0.147  & 0.236 &    0.139  &  -0.24  & -0.272   & -0.167   &  0.158   &  0.056   & 0.127  \\
\#  4 &  -0.092  & 0.214   & \bf{0.437}  &  -0.046  &  0.177  &  0.059  &  -0.22  & -0.213   & -0.177   &  0.174   &  0.009   & 0.229  \\
\#  5 &    0.31  & -0.089  & \bf{0.422} & -0.080   &  0.324  &  0.225  & -0.276  & -0.308   & -0.025   &  0.236   &  0.085   & 0.343  \\
\#  6 &    0.07  & \bf{0.431} &  -0.330  &  0.034 &  0.269  &  0.219  & -0.153  & -0.161   &  0.028   &   0.15   & -0.068   & 0.178  \\
\#  7 &   0.207  & -0.537 & -0.258  & -0.068  &  \bf{0.288}  &  0.124  & -0.324  & -0.302   & -0.307   &  0.198   &  -0.06   & 0.205  \\
\#  8 &   \bf{0.397}  & 0.079 & -0.296  & -0.226  & 0.28  &  0.141  &  -0.17  & -0.171   & -0.223   &  0.051   &  0.028   & 0.215  \\
\#  9 &   0.291  & -0.007  & 0.164   & \bf{0.483} & 0.284  &  0.108  & -0.041  & -0.026   &  0.146   &  0.118   & -0.087   & 0.223  \\

   \hline

\end{tabular}
\caption{AC for a lead time of $\tau=6$, averaged over all periods in the hold-out sample, by CRD. For each CRD, the model with the highest AC is bolded. See Section \ref{alt_methods} for descriptions of each model listed. }
\label{table: tau_6_AC}
\end{table} 

\begin{table}[H]
\footnotesize
\captionsetup{font=footnotesize}
\centering
\centering
\begin{tabular}{p{1cm} p{1cm} p{1cm}  p{1cm}   p{1cm}   p{1cm}   p{1cm}  p{1cm}   p{1cm} p{1cm}  p{1cm}   p{1cm}  p{1cm} }
&  & & & &   \multicolumn{3}{c}{ Model}  \\
\hline
CRD & BA1 & BA2 &BA3 & BA4 & M1 & M2 & M3 & M4 & M5 & M6 & M7  & M8  \\ [1ex] 
\hline
\#  1 &  6.753 & 6.377   & 6.724  & 5.453  &  4.975  & \bf{4.843}  & 5.121  & 5.153   & 5.139   & 9.425   &   5.51   &  6.45  \\
 \#  2 &  4.556  & 4.814  & 4.620  &  4.502   & 2.674  &  \bf{2.59}  & 4.788  & 4.771   & 4.488   & 4.341   &  8.261   & 5.327  \\
 \#  3 &  7.261  & 7.094  & 6.947  &  7.739   & 5.243  & \bf{4.905}  & 6.998  & 7.351   & 5.964   & 5.575   & 13.379   & 7.358  \\
 \#  4 &  6.311  & \bf{3.929}  & 4.960   &  5.726  & 4.044  & 3.994  & 5.214  & 5.279   &  4.88   & 6.706   &  7.433   & 5.797  \\
 \#  5 &  5.486  & 5.364  & 4.438  & 5.233   & 3.291  & \bf{3.126}  & 5.715  & 6.163   & 4.823   & 5.035   &  8.155   & 5.103  \\
\#  6 &  6.406  & 6.711  &  6.241 &  6.484   & 5.058  & \bf{4.795}  & 6.617  & 6.655   & 5.859   & 8.826   & 13.646   & 6.928  \\
 \#  7 &  3.842  & 4.773  & 4.179  & 4.336   & 2.979  & \bf{2.849}  & 4.523  &  4.69   & 4.041   & 3.802   & 10.431   & 4.789  \\
\#  8 &  5.547  & \bf{5.069}  &  6.371  & 6.954  & 5.813  & 5.669  &  6.67  & 6.865   & 5.626   & 6.427   & 10.086   & 7.827  \\
\#  9 &  3.215  &  5.991 & 9.970   & \bf{2.903}  & 3.404  & 3.476  & 4.627  & 4.588   &  3.67   & 5.879   &  7.595   & 4.864  \\
\hline

\end{tabular}
\caption{MSE values for a lead time of $\tau=3$, averaged over all periods in the hold-out sample, by CRD. For each CRD, the model with the lowest MSE is bolded. See Section \ref{alt_methods} for descriptions of each model listed. }
\label{table: tau_3_MSE}
\end{table}  
\begin{table}[H]
\footnotesize
\captionsetup{font=footnotesize}
\centering
\begin{tabular}{p{1cm} p{1cm} p{1cm}  p{1cm}   p{1cm}   p{1cm}   p{1cm}  p{1cm}   p{1cm} p{1cm}  p{1cm}   p{1cm}  p{1cm} }
&  & & & &   \multicolumn{3}{c}{ Model}  \\
\hline
CRD & BA1 & BA2 &BA3 & BA4 & M1 & M2 & M3 & M4 & M5 & M6 & M7  & M8  \\ [1ex] 
\hline
\#  1 &  -0.419  & -0.128  & -0.219 & -0.038   &   0.078  & \bf{0.289}  &  0.059  &  0.045   &  0.082   & -0.026   &  0.056   &  0.002  \\
 \#  2 &  -0.054  &  -0.090   & 0.141 & -0.045  & 0.195  & \bf{0.678}  & -0.184  & -0.126   & -0.145   &  0.122   &  0.086   &  0.012  \\
\#  3 &  -0.317  & -0.438  & -0.222   & -0.327  & 0.282  & \bf{0.391}  &  -0.24  & -0.272   & -0.167   &  0.158   &  0.056   & -0.045  \\
\#  4 &  -0.257   & \bf{0.371}  & -0.049   & -0.234 &  0.255  & 0.367  &  -0.22  & -0.213   & -0.177   &  0.174   &  0.009   & -0.017  \\
\#  5 &  -0.216  & -0.152 & 0.248  & -0.152  &  0.365  & \bf{0.589}  & -0.276  & -0.308   & -0.025   &  0.236   &  0.085   & -0.005  \\
\#  6 &   -0.05  & -0.360  & 0.098   & -0.202  &  0.386  & \bf{0.418}  & -0.153  & -0.161   &  0.028   &   0.15   & -0.068   &   0.14  \\
\#  7 &   0.083  & -0.334   & -0.046   & -0.090  &  0.265  & \bf{0.437}  & -0.324  & -0.302   & -0.307   &  0.198   &  -0.06   & -0.098  \\
\#  8 &   0.011  & 0.237  & -0.036   & -0.203   & \bf{0.245}  & 0.121  &  -0.17  & -0.171   & -0.223   &  0.051   &  0.028   &  0.016  \\
\#  9 &   0.395  & -0.267   & -0.261  &\bf{ 0.557}  &  0.275  & 0.362  & -0.041  & -0.026   &  0.146   &  0.118   & -0.087   &  0.135  \\

\hline

\end{tabular}
\caption{AC for a lead time of $\tau=3$, averaged over all periods in the hold-out sample, by CRD. For each CRD, the model with the highest AC is bolded. See Section \ref{alt_methods} for descriptions of each model listed. }
\label{table: tau_3_AC}
\end{table} 
\begin{table}[H]
\footnotesize
\captionsetup{font=footnotesize}
\centering
\begin{tabular}{p{1cm} p{1cm} p{1cm}  p{1cm}   p{1cm}   p{1cm}   p{1cm}  p{1cm}   p{1cm} p{1cm}  p{1cm}   p{1cm}  p{1cm} }
&  & & & &   \multicolumn{3}{c}{ Model}  \\
\hline
CRD & BA1 & BA2 & BA3 & BA4 & M1 & M2 & M3 & M4 & M5 & M6 & M7  & M8  \\ [1ex] 
\hline
\#  1 &   4.93  & 6.759  & 8.287   &  \bf{4.800}   &   7.99  &  7.83  & 5.121  & 5.153   & 5.139   & 9.425   &   5.51   & 5.316  \\
 \#  2 &   4.45  & 4.637   & 5.920 & 4.473   &  5.816  &  5.48  & 4.788  & 4.771   & 4.488   & \bf{4.341}   &  8.261   & 4.714  \\
  \#  3 &  6.794  &  5.666  & 5.777 & 6.832   &   7.68  & 7.444  & 6.998  & 7.351   & 5.964   & \bf{5.575}   & 13.379   & 5.988  \\
  \#  4 &  5.612  & 4.984   & 5.315 & 5.651   &  6.002  & 6.013  & 5.214  & 5.279   &  4.88   & 6.706   &  7.433   & \bf{4.391}  \\
  \#  5 &  5.237  & 4.767  & 6.629  & 4.821  & 5.832  & 5.624  & 5.715  & 6.163   & 4.823   & 5.035   &  8.155   &  \bf{4.31}  \\
  \#  6 &  5.659  & 6.285  & 7.902 & 6.171  & 5.956  & 5.859  & 6.617  & 6.655   & 5.859   & 8.826   & 13.646   & \bf{4.488}  \\
  \#  7 &   4.03  & 4.172    & 3.273 & 4.372 & 3.642  & 3.672  & 4.523  &  4.69   & 4.041   & 3.802   & 10.431   & \bf{2.983}  \\
  \#  8 &  5.998  & 4.868   & 7.115  & 6.717 & 6.941  & 7.025  &  6.67  & 6.865   & 5.626   & 6.427   & 10.086   & \bf{4.549}  \\
  \#  9 &  2.743  & 4.446    & 5.569  & \bf{2.212}   &  5.533  & 5.485  & 4.627  & 4.588   &  3.67   & 5.879   &  7.595   & 3.428  \\
    \hline
\end{tabular}
\caption{MSE values for a lead time of $\tau=1$, averaged over all periods in the hold-out sample, by CRD. For each CRD, the model with the lowest MSE is bolded. See Section \ref{alt_methods} for descriptions of each model listed. }
\label{table: tau_1_MSE}
\end{table}
\begin{table}[H]
\footnotesize
\centering
\captionsetup{font=footnotesize}
\begin{tabular}{p{1cm} p{1cm} p{1cm}  p{1cm}   p{1cm}   p{1cm}   p{1cm}  p{1cm}   p{1cm} p{1cm}  p{1cm}   p{1cm}  p{1cm} }
&  & & & &   \multicolumn{3}{c}{ Model}  \\
\hline
CRD & BA1 & BA2 &BA3 & BA4 & M1 & M2 & M3 & M4 & M5 & M6 & M7  & M8  \\ [1ex] 
\hline
  \#  1 &   0.233  &   -0.368  & -0.472    &   \bf{0.301} &   0.087  & -0.409  &  0.059  &  0.045   &  0.082   & -0.026   &  0.056   & 0.087  \\
 \#  2 &   0.038  & 0.014  & -0.422    &  -0.015 &  0.243  &   0.06  & -0.184  & -0.126   & -0.145   &  0.122   &  0.086   & \bf{0.245}  \\
 \#  3 &  -0.214  & 0.142  & 0.115  &  -0.075   &  \bf{0.296}  & -0.043  &  -0.24  & -0.272   & -0.167   &  0.158   &  0.056   & 0.285  \\
 \#  4 &   0.024  & 0.057  & 0.006  & -0.033  &  \bf{0.217}  & -0.075  &  -0.22  & -0.213   & -0.177   &  0.174   &  0.009   &  0.18  \\
 \#  5 &   -0.23  & 0.131 &  -0.272  & 0.032  &  \bf{0.346}  &  0.235  & -0.276  & -0.308   & -0.025   &  0.236   &  0.085   & 0.316  \\
 \#  6 &   0.164  & -0.019  & -0.457 & -0.135    &  \bf{0.341}  &  0.311  & -0.153  & -0.161   &  0.028   &   0.15   & -0.068   & 0.282  \\
 \#  7 &   0.024  &  -0.123 & \bf{0.288} & -0.051  & 0.259  &  0.197  & -0.324  & -0.302   & -0.307   &  0.198   &  -0.06   & 0.139  \\
 \#  8 &  -0.187  & \bf{0.298}  & -0.301 & -0.187  & 0.221  &   0.07  &  -0.17  & -0.171   & -0.223   &  0.051   &  0.028   & 0.218  \\
 \#  9 &   0.542  & -0.142   & -0.418  & \bf{0.747}  &  0.298  &  0.218  & -0.041  & -0.026   &  0.146   &  0.118   & -0.087   & 0.248  \\
   \hline
\end{tabular}
\caption{AC for a lead time of $\tau=1$, averaged over all periods in the hold-out sample, by CRD. For each CRD, the model with the highest AC is bolded. See Section \ref{alt_methods} for descriptions of each model listed. }
\label{table: tau_1_AC}
\end{table} 
\end{document}